\documentclass{article}
\usepackage{bookstyle}
\usepackage{graphicx}

\begin{document}

\author{Bertrand Reulet}
\address{Departments of Applied Physics and Physics\\
 Yale University,  New Haven CT 06520-8284, USA\\
Laboratoire de Physique des Solides, UMR8502, b\^at 510\\
Universit\'e Paris-Sud, 91405 Orsay, France\\}
\title{Higher moments of noise}
%
\frontmatter
\maketitle
\mainmatter

\section{Introduction}

Transport studies provide a powerful tool for investigating electronic properties of a conductor.  The $I(V)$ characteristic (or the differential resistance $R_{diff}=dV/dI$) contains partial information on the mechanisms responsible for conduction.  A much more complete description of transport in the steady state, and further information on the conduction mechanisms, is given by the probability distribution of the current $P$, which describes both the dc current $I(V)$ and the fluctuations. Indeed, even with a fixed voltage $V$ applied, $I(t)$ fluctuates, due to
the discreteness of the charge carriers, the probabilistic character of scattering and the fluctuations of the population of energy levels at finite temperature $T$ \cite{BuBlan}.

The current fluctuations are characterized by the moments of the probability distribution $P$ of order two and higher. Experimentally, the average over $P$ is obtained by time averaging. Thus, the average current is the dc current $I=\left<I(t)\right>$, where $\left<.\right>$ denotes time average. The second moment (the variance) of $P$, $\left<i^2\right>$, measures the amplitude of the current fluctuations, with $i(t)=I(t)-I$. The third moment $\left<i^3\right>$ (the skewness) measures the asymmetry of the fluctuations.
Gaussian noise $P(i)\propto{\rm exp}(-\alpha i^2)$ is symmetric, so it has no third moment.
The existence of the third moment is related to the breaking of time reversal symmetry by the dc current; at zero bias, $I=0$ and positive and negative current fluctuations are equivalent, so $\left<i^3\right>=0$.

In this article we present simple approaches to the calculation of $P(i)$ in a tunnel junction, and to the effect of the environment on noise measurements in terms of the modification of $P$. We do not provide rigorous calculations, but simple considerations that bear the essential ingredients of the phenomena. We also discuss the effect of a finite measurement bandwidth. We report experimental results of the first measurement of the third moment of voltage fluctuations in tunnel junctions, from room temperature down to $50$mK. Then we discuss extensions of that measurements to finite frequencies and to the study of other systems. We show the first data of the third moment in the regime where the frequency is larger than the temperature. Finally we discuss a new quantity, the "noise thermal impedance", which links the second and third moment.

\section{The probability distribution $P(i)$}

\subsection{A simple model for a tunnel junction}

Let us consider a single channel tunnel junction of transmission probability $t$. For $t\ll1$, the tunneling events are rare and well separated. Thus one can consider a time $\tau$ small enough such that there is at most one event during $\tau$. The transport properties of the junction are given by the rates $\Gamma_+$ and $\Gamma_-$ at which the electrons cross the barrier from left to right or vice-versa. One electron crossing the junction during a time $\tau$ corresponds to a current pulse during $\tau$  of average intensity $\bar{\imath}=e/\tau$. The probability for $n$ electrons to cross the barrier during $\tau$ from left to right, giving rise to a current $n\bar{\imath}$ is:
\begin{equation}
\left\{
\begin{array}{l}
P(0) =  1-(\Gamma_++\Gamma_-)\tau\\
P(\pm\bar{\imath})  =  \Gamma_\pm\tau\\
P(n\bar{\imath})  =0\;\; \mathrm{ for }\;\; n>1\\
\end{array}
\right.
\end{equation}
The quantum mechanics enters in the calculation of the rates whereas the statistical mechanics of the junction is given by the probability $P(i)$.  We deduce the $p^{th}$ moment of the distribution of the current:
\begin{equation}
\left<I^p\right>=\sum_{n=\pm1}P(n\bar{\imath})(n\bar{\imath})^p=\bar{\imath}^p(\Gamma_++(-1)^p\Gamma_-)\tau
\end{equation}
Thus, all the odd moments are proportional to the dc current $I=\left<I\right>=e(\Gamma_+-\Gamma_-)$, and all the even moments are proportional to the second one $\left<I^2\right>=e^2(\Gamma_++\Gamma_-)\tau$.
The value of $\Gamma_\pm$ are determined by Ohm's law $I=e(\Gamma_+-\Gamma_-)=GV$ and the detailed balance $\Gamma_+/\Gamma_-=\exp(eV/k_BT)$. In particular at zero temperature and $V>0$, $\Gamma_-=0$, which gives $\left<I^2\right>=eI\tau^{-1}$

One generally considers the moments of the \textit{current fluctuations}, $M_p=\left<i^p\right>$ with $i(t)=I(t)-\left<I\right>$. One has for the first moments:
\begin{equation}
\nonumber
\begin{array}{l}
M_2=\left<i^2\right>=\left<I^2\right>-I^2\\
M_3=\left<i^3\right>=\left<I^3\right>-3I\left<I^2\right>+2I^3
\end{array}
\end{equation}
Since $\left<I^p\right>\propto t$ for all $p$, one has for a tunnel junction with $t\ll1$, $\left<i^p\right>\simeq\left<I^p\right>$ to leading order in $t$. In particular, $M_3=e^2I\tau^{-2}$. This result is valid even for the multichannel case and after integrating over energy \cite{LevitovMath}. It is remarkable that $M_3$ is totally temperature independent, in contrast with $M_2$, for which the fluctuation-dissipation theorem implies $M_2\propto T$ at equilibrium. As expected, $M_3$ is an odd function of the dc current and is zero for $V=0$.

One usually also defines the \emph{cumulants} $C_p$ of current fluctuations (often noted $\left<\left< I^p\right>\right>$). They are related to the moments through the Fourier transform $\chi(q)=\sum_nP(n\bar\imath)\exp iqn$ of $P$. The series expansion of $\chi$ gives the moments, whereas
 the series expansion of $\ln\chi$ give the cumulants. For a Gaussian distribution, $\ln\chi$ is a second degree polynomial, and thus all the cumulants of order $\geq3$ are zero. The cumulants measure the non-gaussian part of the noise. The $C_p$ are linear combinations of products of the $M_p$. For example,
\begin{equation}
\begin{array}{l}
C_3=M_3\\
C_4=M_4-3M_2^2
\end{array}
\end{equation}

\subsection{The noise in Fourier space}

In the previous paragraph we used a simple model to calculate the probability of current fluctuations. Experimentally, the averaging procedure is usually done by integrating over time the desired quantity. However, the fluctuating current $i(t)$ contains Fourier components up to very high frequency which are usually not accessible experimentally. Thus, one rather measures the \textit{spectral density} of the fluctuations around a certain frequencies.
We introduce the spectral densities associated with the $p^{\mathrm{th}}$ moment of the current fluctuations $S_{I^p}$ , expressed in A$^p$/Hz$^{p-1}$. $S_{I^p}$ depends on $p-1$ frequencies $f_1\dots f_{p-1}$. However, it is convenient to express $S_{I^p}$ as a function of $p$ frequencies such that the sum of all the frequencies is zero. Introducing the Fourier components $i(f)$ of the current, one has, for a classical current:
\begin{equation}
S_{I^p}(f_1,\dots,f_{p-1})=\left<i(f_1)\dots i(f_p)\right>\delta(f_1+\dots f_p)
\end{equation}
In quantum mechanics, the current operators taken at different times do not commute; they also do not in Fourier space, and the question of how the operators have to be ordered is crucial \cite{LevitovMath,Chtche}.

In the case of the second moment, one has $S_{I^2}(f)=\left<i(f)i(-f)\right>=\left<|i(f)|^2\right>$. It measures the power emitted by the sample at the frequency $f$ within a bandwidth of 1Hz. This is what a spectrum analyzer measures. Comparing this expression with the one we have calculated for $\left<\delta I^2\right>$, one sees that $\tau^{-1}$ roughly represents the full bandwidth of the current fluctuations.

Experimentally, the current emitted by the sample runs through a series of cables, filters and amplifiers before being detected (this can be avoided by an on-chip detection \cite{richard}). Thus, the measured quantity is a filtered current $j(f)=i(f)g(f)$ where $g(f)$ describes the filter function of the detection.

One is often interested in the total power emitted by the sample in a certain bandwidth. This is obtained by measuring the DC voltage after squaring $j(t)$, i.e. $\int j^2(t)dt$. This quantity is related to $S_{I^2}$ through:
\begin{equation}
\begin{array}{ll}
\left<j^2\right> & =\int\!\!j^2(t)dt=
\int\int_{-\infty}^{+\infty}
g(f_1)g(f_2)\left<i(f_1)i(f_2)\right>\delta(f_1+f_2)\\
& =\int |g(f)|^2S_{I^2}(f)df
\end{array}
\end{equation}
It is remarkable that the frequency-dependent phase shift introduced by $g(f)$ has no influence, in agreement with the fact that $S_{I^2}$ has the meaning of a power.
If the detection bandwidth extend from $F_1$ to $F_2$, i.e., $g(f)=1$ for $F_1<|f|<F_2$ and $g(f)=0$ otherwise, and if $S_{I^2}$ is frequency independent between $F_1$ and $F_2$, then the total noise is given by $\left<j^2\right>=2S_{I^2}(F_2-F_1)$.

\begin{figure}
\includegraphics[width= 0.7\columnwidth]{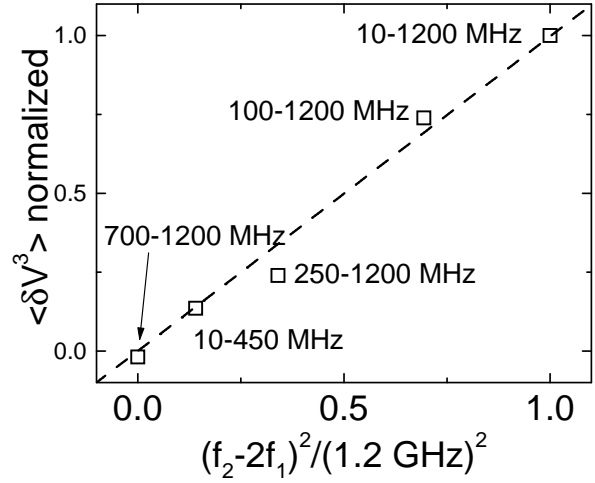}
\caption{Effect of finite bandwidth on the measurement of $\left<\delta V^3\right>$. Each point corresponds to a different value of the frequencies $F_1$ and $F_2$, as indicated in the plot. The data shown here correspond to sample B at $T=77$ K. (from Ref.\cite{S3nous}).}
\label{bandwidth}
\end{figure}

Let us consider now $S_{I^3}$, which depends on two frequencies $f_1$ and $f_2$. The third moment of the measured current $j(t)$ is given by:
\begin{equation}
\left<j^3\right>=\!\!\int\!\! j^3(t)dt=
\int\!\!\!\!\int\!\!\!\!\int g(f_1)g(f_2)g(f_3) S_{I^3}(f_1,f_2)\delta(f_1+f_2+f_3) 
\end{equation}
We see that now the phase of $g(f)$ matters. More precisely, $S_{I^3}$ measures how three Fourier components of the current can beat together to give a non-zero result, i.e., it measures the phase correlations between these three Fourier components. With the same hypothesis as before, for a detection between $F_1$ and $F_2$, one has now $\left<j^3\right>=3S_{I^3}(F_2-2F_1)^2$ if $F_2>2F_1$ and $\left<j^3\right>=0$ otherwise. This shows how important it is to make a broadband measurement. This unusual dependence of the result on $F_1$ and $F_2$ comes from the fact that the lowest frequency being the sum of two others is $2F_1$ whereas the maximum frequency one can subtract to that in order to have a DC signal is $F_2$.
 As we show below, we have experimentally confirmed this unusual dependence of $\left<j^3\right>$ on $F_1$ and $F_2$, see fig. \ref{bandwidth}.

\subsection{Consequences}

Each moment of the distribution is affected in a different way by the finite bandwidth of the measurement. As a consequence, even if the case where the moments are supposed to be frequency-independent, the probability $P(i)$ measured within a finite bandwidth $B$ depends on $B$. In particular, the higher moments, which are more sensitive to rare events (current spikes) are washed out by the finite bandwidth, since the spikes are broadened by the filtering. In the case where the moments are frequency-dependent (like the diffusive wire, \cite{Pilgram_diff}), the notion of counting statistics itself has to be taken with care.

\section{Effect of the environment}

Until now we have considered the bias voltage $V$ and the temperature $T$ to be fixed, time-independent, external parameters. In practice, it is very hard to perfectly voltage-bias a sample at any frequency. The temperature of the sample is generally fixed by a connection to reservoirs, and thus the temperature is fixed only at the ends of the sample. If $V$ or $T$ fluctuate, the probability $P(i)$ is modified. Let us call $P(i;V,T)$ the probability distribution of the current fluctuations around the dc current $I$ when the sample is perfectly biased at voltage $V$ and kept at fixed temperature $T$ (as considered before), and $\tilde{P}(i)$ the probability distribution in the presence of an environment. $R$ is the sample's resistance, taken to be independent of $T$ and $V$.

\subsection{Imperfect voltage bias}

If the sample is biased by a voltage $V$ through an impedance $Z$, the dc voltage across the sample is $V_s=VR/(R+Z)$. However, the current fluctuations in the sample flowing through the external impedance induce voltage fluctuations across the sample, given by:
\begin{equation}
\delta V_s(t)=-\int_{-\infty}^{+\infty} Z(f)i(f)e^{2i\pi ft}df
\end{equation}
Consequently, the probability distribution of the fluctuations is modified. This can be taken into account if the fluctuations are slow enough that the distribution $P(i)$ follows the voltage fluctuations. Under this assumption one has:
\begin{equation}
\tilde{P}(i)=P(i;V_s+\delta V_s,T)
\end{equation}
Supposing that the fluctuations are small ($\delta V_s\ll V_s,k_BT$), one can Taylor expand $P_0$ in:
\begin{equation}
\tilde{P}(i)\simeq P(i;V_s,T)+\delta V_s\frac{\partial P}{\partial V_s}+...
\end{equation}
One deduces the moments of the distribution (to first order in $\delta V_s$) for a frequency-independent $Z$:
\begin{equation}
\tilde{M}_n(V,T)=M_n(V_s,T)-Z\frac{\partial M_{n+1}(V_s,T)}{\partial V_s}
\end{equation}
This equations shows that environmental correction to the moment of order $n$ is related to the next moment of the sample perfectly voltage biased. It is a simplified version of the relation derived in refs. \cite{Saleur,Zaikin_cumulants}.
Let us now apply the previous relation to the first moments.

\subsubsection{dc current: dynamical Coulomb blockade}

For $n=1$ one gets a correction to the dc current given by:
\begin{equation}
\left< i\right>=-\int_{-\infty}^{+\infty} Z(f)\frac{\partial M_2(f)}{\partial V_s}df
\end{equation}
This is nothing but the environmental Coulomb blockade (within a factor 2) \cite{Ingold,yeyati}. The bandwidth involved in $M_2$ is the intrinsic bandwidth of the sample, limited by the RC time,  and not the detection bandwidth.

\subsubsection{The second moment}

Since the intrinsic third moment of a tunnel junction is linear in the applied voltage, to lowest order the imperfect voltage bias affects the second moment only by a constant term. There are however second order corrections \cite{gefen_tunnel}.

\subsubsection{The third moment}

Similarly one obtains \cite{Kindermann}:
\begin{equation}
\tilde{M}_3=M_3+3ZM_2\frac{\partial M_2}{\partial V_s}
\end{equation}
It is clear that the environmental correction to the third moment and the dynamical Coulomb blockade share the same physical origin, i.e. electron-electron interactions. However, since the third cumulant is a small quantity, the corrections can be as large as the intrinsic contribution, especially in a low impedance sample, as the one we have measured.

\subsubsection{Effect of an external fluctuating voltage}

\begin{figure}
\includegraphics[width= 0.8\columnwidth]{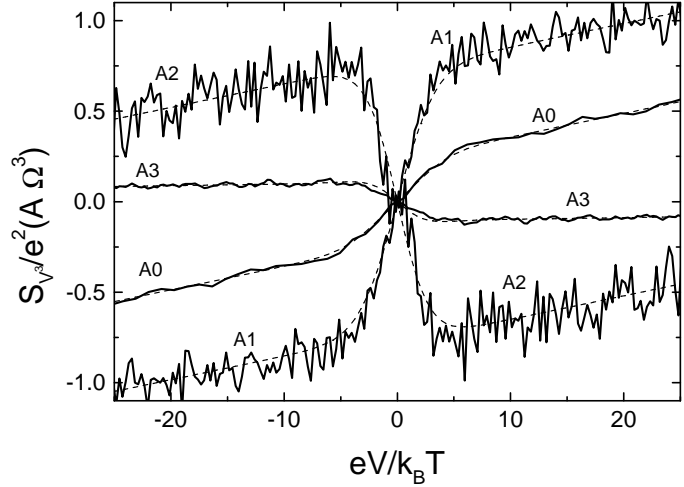}
\caption{Measurement of the spectral density of the third moment of voltage fluctuations, $S_{V^3}(eV/k_BT)$ for sample A at T=4.2 K (solid lines). A0: no ac excitation (same as Fig. \ref{samplea}). A1: with an ac excitation at frequency $\Omega/2\pi$ such that $\cos2\Omega\Delta t=+1$; A2: $\cos2\Omega\Delta t=-1$; A3: no ac excitation but a $63\;\Omega$ resistor in parallel with the sample. The dashed lines corresponds to fits with Eq. (\ref{eqenv}). (from Ref. \cite{S3nous}).}
\label{env}
\end{figure}

We consider in this paragraph the effect of an external source of noise $i_0(t)$ in parallel with the external impedance $Z$. This noise could be due to the Johnson noise of $Z$, to backaction of the current detector (as the current noise of an amplifier), or an applied signal. The total current contains now $i_0$, and its moments involve correlations between $i_0$ and $i$. For the third moment, an additional term appears due to $i_0$: $\left<i_0i^2\right>\simeq Z\left<i_0^2\right>(\partial M_2/\partial V_s)$. Our measurements verified the two effects of the environment (feedback and external noise) on the third moment, see fig. \ref{env}.

\subsubsection{Voltage vs. current fluctuations}

Instead of applying a voltage and measure a current, one often prefers to apply a current and measure voltage fluctuations across the sample. The first two moments of $V_S$ and $i$ are related through:
\begin{equation}
\nonumber
\begin{array}{l}
V_S=R_DI\\
\left< \delta V_S^2\right>=R_D^2 \left(M_2+\left<i_0^2\right>\right)
\end{array}
\end{equation}
with $R_D=RZ/(R+Z)$. The situation is different for the third moment, for which we have\cite{Kindermann2,S3nous}:
\begin{equation}
\left<\delta V_S^3\right>=-R_D^3M_3+
3R_D^4\left< i_0^2\right>\frac{\partial M_2}{\partial V_S}
+3R_D^4M_2^0\frac{\partial M_2}{\partial V_S}
\label{eqenv}
\end{equation}
We have confirmed this relation experimentally, see fig.\ref{env}.

\subsection{Imperfect thermalization}

Let us suppose now that the sample is perfectly voltage biased, but that its temperature $T_s$ can fluctuate, because there is a finite thermal impedance $G_{th}$ between the sample and the reservoir. If the sample is biased at a finite voltage $V$, its average temperature might be different from the reservoir's temperature. Since the current flowing through the sample fluctuates, the Joule power $P_J$ dissipated in the sample fluctuates as well, which induces temperature fluctuations. One has: $\delta T_s=G_{th}^{-1}\delta P_J=G_{th}^{-1}iV$. Since the probability distribution $P(i)$ depends on temperature, the temperature fluctuations modify in turn the current fluctuations. This \emph{thermal} feedback is similar to the \emph{electronic} feedback of the previous section. Thus, one has:
\begin{equation}
\tilde{P}(i)=P(i;V,T_s+\delta T_s)\simeq P(i;V,T_s)+\delta T_s\frac{\partial P}{\partial T_s}+...
\end{equation}
This results in the following equation for the moments:
\begin{equation}
\tilde{M}_n(V,T)=M_n(V,T_s)+G_{th}^{-1}V\frac{\partial M_{n+1}}{\partial T_s}
\end{equation}
Note that $G_{th}$ is in fact complex, temperature- and frequency dependent.

Similarly to the case of imperfect voltage bias, the dc current and the third moment are affect by this feedback. One obtains a relative correction to the conductance $\delta G/G\propto k_B/C$ with $C$ the heat capacitance of the sample. This correction is small, but diverges at low temperature. To our knowledge, this had never been predicted before.

For the third moment one obtains:
\begin{equation}
\tilde{M}_3=M_3+6VG_{th}^{-1}M_2\frac{\partial M_2}{\partial T}
\end{equation}
In the case of a diffusive wire whose length is much longer than the electron-electron inelastic length, a local temperature can be defined. We designate $T$ the (voltage-dependent) average temperature of the sample. In the absence of electron-phonon interactions, the electrical and thermal conductances are related through the Wiedemann-Franz law, $G_{th}\propto GT$ ($G=R^{-1}$), from which we deduce:
\begin{equation}
\tilde{M}_3=M_3+\alpha e^2IB^2
\end{equation}
with $\alpha$ a numerical coefficient.
This result corresponds to the calculation of ref. \cite{Pilgram_diff} in the hot electron regime. We understand that at frequencies larger than the inverse of the diffusion time $\tau_D$, the thermal conductance drops, and so does the thermal correction. The result of ref. \cite{Pilgram_diff}, that the third moment vanishes at such frequencies, seems to imply that $M_3=0$. It might be an artifact of our oversimplified calculation. Here $B$ has the meaning of the thermal bandwidth, $B\sim\tau_D^{-1}$.

Another interesting case is the SNS structure (a normal metal wire between two superconducting reservoirs), for which the cooling is due to electron-phonon interaction and not electron out-diffusion. This out-diffusion is suppressed exponentially at low temperature due to the superconducting gap of the reservoirs. The vanishing electron-phonon thermal conductance should lead to a divergence of the spectral density of the third moment, but the bandwidth also shrinks with lowering $T$.

\section{Principle of the experiment}

\begin{figure}
\includegraphics[width= 0.8\columnwidth]{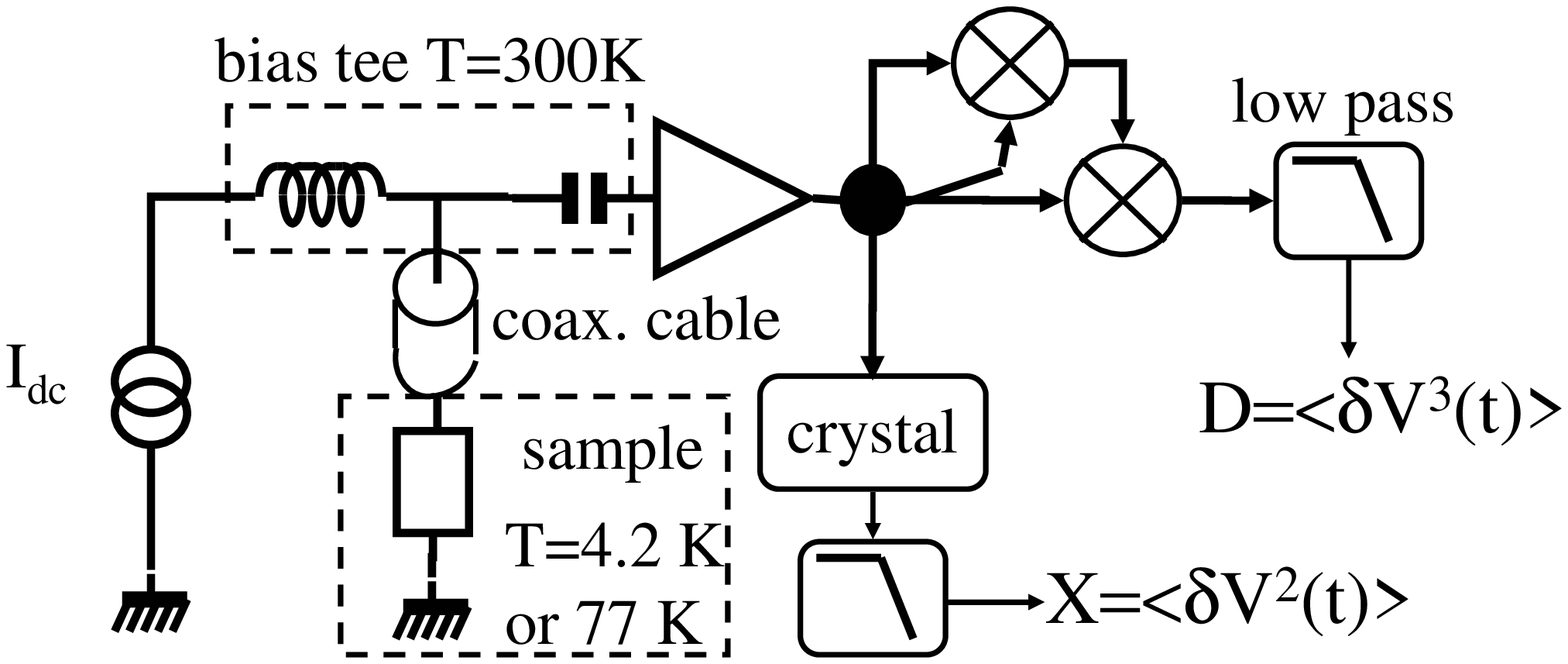}
\caption{Schematics of the experimental setup. (from Ref. \cite{S3nous})}
\label{setup}
\end{figure}

We will not discuss the details of the experiment performed to measure the third moment of the voltage fluctuations in tunnel junctions. Those details can be found in refs. \cite{S3nous,SPIE}.

\subsection{Possible methods}

We present three methods that could be used to measure the non-Gaussian part of the noise. First, the simplest idea is to digitize in real time the voltage across the sample, and make histograms of the values. This method is very direct, but suffers from its limitations in bandwidth $B$. It is very difficult to acquire data with enough dynamics very fast and to treat them in real time. Since $M_3\propto B^2$, the small bandwidth severely limits this method. No such measurement has been reported yet.

A second point of view is to put the priority on the bandwidth. That is the method we have chosen. Then it is very hard to treat the signal digitally, and we use analog mixers to compute the third power of $i(t)$, and average with the help of a low pass filter. We have implemented this method with a bandwidth of 1GHz. This method is versatile, it works at any temperature and for many different kind of samples; the main limitation comes from the necessity to have a sample's resistance close to $50\Omega$ to ensure good coupling with the microwave circuits. The drawback of the method is the care needed to separate the real signal from any non-linearity due to the mixers and amplifiers.

Finally, another possibility, which offers huge bandwidth and great sensitivity is to couple the sample to an on-chip mesoscopic detector. This has been successfully realized to measure $S_{V^2}$ \cite{richard}. One could even have access to the full statistics of the current\cite{tobiska}. The drawback of this method is the theoretical difficulty of extracting the behavior of the sample while it is strongly coupled to another mesoscopic system\cite{finland}.

\subsection{Experimental setup}

We have measured the third moment of the voltage fluctuations across a tunnel junction, by measuring $\delta V^3(t)$ in real time (see Fig. \ref{setup}). For simplicity, we note $V$ and $\delta V$ are the dc voltage and the voltage fluctuations \emph{across the sample}.
The sample is dc current biased through a bias tee. The noise emitted by the sample is coupled out to an rf amplifier through a capacitor so only the ac part of the current is amplified. The resistance of the sample is close to $50\;\Omega$, and thus is well matched to the coaxial cable and amplifier. After amplification at room temperature the signal is separated into four equal branches, each of which carries a signal proportional to $\delta V(t)$. A mixer multiplies two of the branches, giving $\delta V^2(t)$; a second mixer multiplies this result with another branch. The output of the second mixer, $\delta V^3(t)$, is then low pass filtered, to give a signal $D$ proportional to $S_{V^3}$, where the constant of proportionality depends on mixer gains and frequency bandwidth. The last branch is connected to a square-law crystal detector, which produces a voltage $X$ proportional to the the rf power it receives: the noise of the sample $\left<\delta V^2\right>$ plus the noise of the amplifiers. This detection scheme has the advantage of the large bandwidth it provides ($\sim1$ GHz), which is crucial for the measurement. Due to the imperfections of the mixers, $D$ contains some contribution of $X$. Those contributions, even in $I$, are removed by calculating $(D(I)-D(-I))\propto S_{V^3}$.

\section{Experimental results}

\subsection{Third moment vs. voltage and temperature}

\begin{figure}
\includegraphics[width= 0.8\columnwidth]{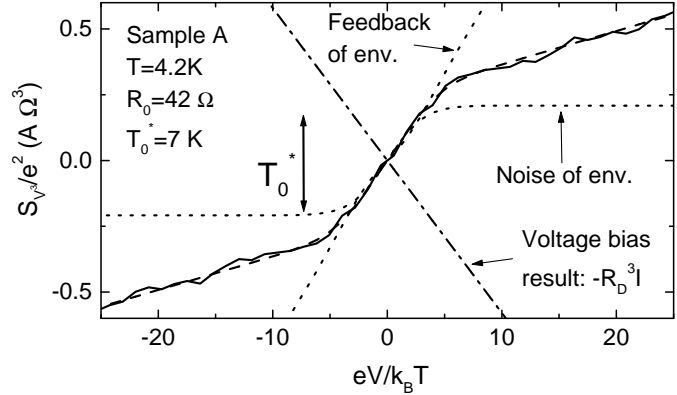}
\caption{Measurement of $S_{V^3}(eV/k_BT)$ for sample A (solid line). The dashed line corresponds to the best fit with Eq. (\ref{eqenv}). The dash dotted line corresponds to the perfect bias voltage contribution and the dotted lines to the effect of the environment. (from Ref. \cite{SPIE})}
\label{samplea}
\end{figure}

\begin{figure}
\includegraphics[width= 0.8\columnwidth]{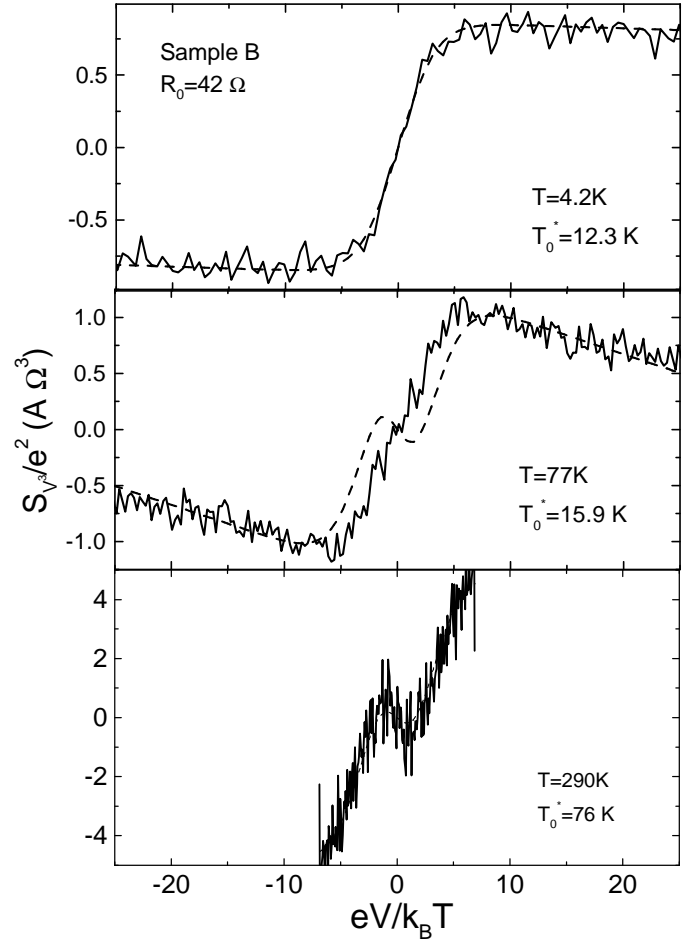}
\caption{Measurement of $S_{V^3}(eV/k_BT)$ for sample B (solid lines). The dashed lines corresponds to the best fit with Eq. (\ref{eqenv}). (from Ref. \cite{SPIE}).}
\label{sampleb}
\end{figure}

$S_{V^3}(eV/k_BT)$ for sample A at $T=4.2$K is shown in Fig. \ref{samplea}; these data were averaged for 12 days.
$S_{V^3}(eV/k_BT)$ for sample B at $T=4.2$K (top), $T=77$K (middle) and $T=290$K (bottom) is shown in Fig. \ref{sampleb}. The averaging time for each trace was 16 hours.
These results are clearly different from the voltage bias result (the dash-dotted line in Fig. \ref{samplea}). However, all our data are very well fitted by Eq. (\ref{eqenv}) which takes into account the effect of the environment (see the dash lines of Fig. \ref{samplea} and \ref{sampleb}). The environment of the sample is made of the amplifier, the bias tee, the coaxial cable ($\sim2$m long except at room temperature, where it is very short) and the sample holder. It is characterized by its impedance $Z$, that we suppose is real and frequency-independent (i.e., we model it by a resistor $R_0$ of the order of $50\Omega$), and a noise temperature $T_0^*$ (the latter does not correspond to the real noise temperature of the environment, see below). Figs. \ref{samplea} and \ref{sampleb} show the best fits to the theory, Eq. (\ref{eqenv}), for all our data. The four curves lead to $R_0=42\;\Omega$, a very reasonable value for microwave components, and in agreement with the fact that
the electromagnetic environment was identical for the two samples.

We have measured directly the noise emitted by the room temperature amplifier; we find $T_0\sim100$ K. This is in disagreement with the parameters $T_0^*$ deduced from the fit of the data, but is well explained by the finite propagation time along the coaxial cable between the sample and the amplifier.

\subsection{Effect of the detection bandwidth}

A powerful check that $D$ really measures $S_{V^3}$ is given by varying the  bandwidth. The scaling of $S_{V^3}$ with $F_1$ and $F_2$ ($S_{V^3}\propto(F_2-2F_1)^2$ if $F_2>2F_1$ and 0 otherwise) is characteristic of the measurement of a third order moment. We do not know any experimental artifact that has such behavior.
 $F_1$ and $F_2$ are varied by inserting filters before the splitter. As can be seen in Fig. \ref{bandwidth}, our measurement follows the dependence on $(F_2-2F_1)^2$, which cannot be cast into a function of $(F_2-F_1)$.
Each point on the curve of Fig. \ref{bandwidth} corresponds to a full $\left<\delta V^3(I)\right>$ measurement (see figures \ref{samplea} and \ref{sampleb}).

\section{Effect of the environment}

In order to demonstrate more explicitly the influence of the environment on $S_{V^3}$ we have modified the parameters $T_0$ and $R_0$ of the environment and measured the effect on $S_{V^3}$.

$T_0$ is a measure of the current fluctuations emitted by the environment towards the sample. Its influence on $S_{V^3}$ is through the correlator $\left<i^2i_0\right>$. This correlator does not require $i_0$ to be a randomly fluctuating quantity in order not to vanish. So we can modify it by adding a signal $A\sin \Omega t$ to $i_0$ (with $\Omega$ within the detection bandwidth). That way we have been able to modify $T_0^*$ without changing $R_0$, as shown on Fig. \ref{env}. The current correlator involved in $T_0^*$ contains interference between the current sent to the sample and what comes back after $2\Delta t$, with $\Delta t $ the propagation time along the coax cable between the source and the sample. This correlator$\sim\left<i_0(t)i_0(t-2\Delta t)\right>$ oscillates vs. $\Delta t$ like $\cos2\Omega\Delta t$, and thus one can enhance (curve A1 as compared to A0 in Fig. \ref{env}) or decrease $T_0^*$, and even make it negative (Fig. \ref{env}, A2). The curves A0--A2 are all parallel at high voltage, as expected, since the impedance of the environment remains unchanged; $R_0=42\;\Omega$ is the same for the fit of the three curves.

Second, by adding a $63\;\Omega$ resistor in parallel with the sample (without the added ac excitation) we have been able to change the resistance of the environment $R_0$, and thus the high voltage slope of $S_{V^3}$. The fit of curve A3 gives $R_0=24.8\;\Omega$, in excellent agreement with the expected value of $25.2\;\Omega$ ($63\;\Omega$ in parallel with $42\;\Omega$). The apparent negative $T_0^*$ comes from the negative sign of the reflection coefficient of a wave on the sample in parallel with the extra resistor.

\section{Perspectives}

\subsection{Quantum regime}

We have chosen to discuss the statistics of the current fluctuations with a classical point of view. This seems enough to explain the properties we have shown until now. However, there are situations that need to be treated quantum mechanically, in particular, when the frequency is greater than voltage and/or temperature. It has been calculated that the third cumulant for a tunnel junction should be completely frequency independent \cite{Zaikin_freq}. This is in sharp contrast with the fact that the second moment $S_{I^2}$ is different for $\hbar\omega>eV$, due to the fact than no photon of energy greater than $eV$ can be emitted. A picture of the third moment (and higher) in terms of photons is still missing. The effect of the environment, which involves $S_{I^2}$, might be also different at high frequency. In particular, the distinction between emission and absorption might be relevant. For example, the zero point fluctuation of the voltage might modulate the noise, but not be detected, and thus some of the effect of the environment might vanish.

\begin{figure}
\includegraphics[width= 0.9\columnwidth]{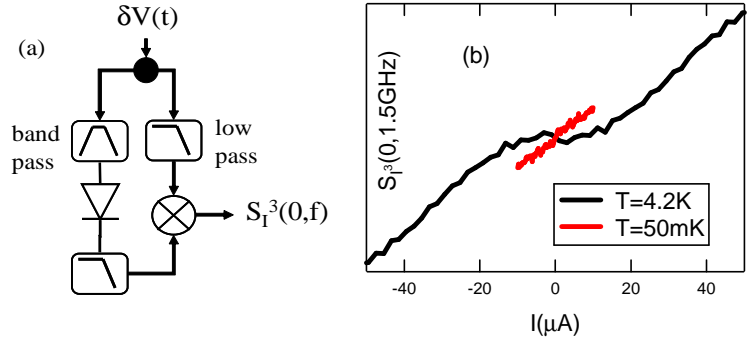}
\caption{(a) Setup for the measurement of $S_{V^3}(0,\bar f)$. The sample, bias tee and  amplifiers have been omitted, see Fig. 1. (b) Measurement of $S_{V^3}(0,\bar f\sim1.5$GHz) on sample A at $T=4.2$K and $T=50$mK.}
\label{fighf}
\end{figure}

We illustrate this discussion by the first measurement of the third moment of voltage fluctuations $S_{V^3}(0,f)$ across a tunnel junction (sample A) at finite frequency $f$ with $hf>k_BT$.
In order to perform such a measurement we have constructed the setup of Fig. \ref{fighf}a. The signal $\delta V(t)$ is split into two frequency bands, LF=$]0,f_3]$ (dc excluded) and HF=$[f_1,f_2]$. The voltage is squared in the HF band (left branch of Fig. 3a) with a high speed tunnel diode, then low-pass filtered with a cutoff $f_3$. Thus one has products of the form $i(f)i(-f')$ at the end of the HF branch. This result is multiplied by a mixer to the LF branch, then low-pass filtered to get a dc signal. This signal corresponds to $S_{V^3}(0,\bar f)$ if $(f_2-f_1)$ and $f_3$ are small enough, with $\bar f\sim(f_1+f_2)/2$. $f_1$ and $f_2$ can be varied by changing the various filters.

\begin{figure}
\includegraphics[width= 0.8\columnwidth]{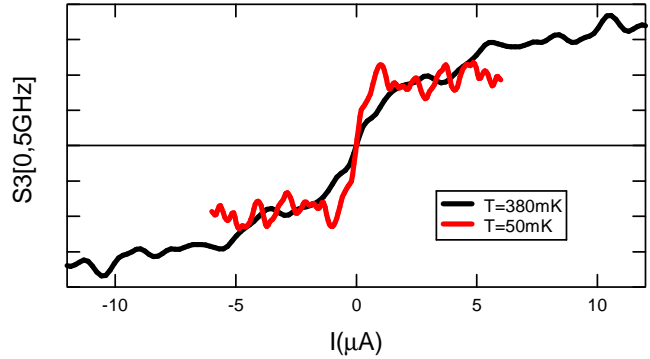}
\caption{(a) Measurement of $S_{V^3}(0,\bar f\sim5$GHz) on sample A at $T=380$mK (i.e., $h\bar f/k_BT\sim0.6$ and $T=50$mK (i.e., $h\bar f/k_BT\sim5$). }
\label{fig_S3q}
\end{figure}

The measurement of sample A at $T=4.2$K and $T=50$mK with the new setup operating in the LF=10-200MHz and HF=1-2.4GHz bandwidths is presented in Fig. \ref{fighf}b. We find subtle but interpretable new results. The fit of these data with eq. (\ref{eqenv}) leads to $R_0=40\Omega$ and $T_0^*(T=4.2K)=-0.4K$. The slope of $S_{V^3}$ at high voltage is found to be temperature independent, like it is between 300K and 4.2K (see Fig. \ref{sampleb}). The negative $T_0^*$ comes from the Johnson noise of the $12\Omega$ contact resistance. The current fluctuations emitted by the contact result in currents of opposite signs running through the sample and the amplifier. As a consequence, the contact contributes to $T_0^*$ with a negative sign. Since we use in this new setup a cryogenic amplifier with low noise temperature $T_0$, and since the Johnson noise of the contact is not affected by the propagation time, its contribution dominates $T_0^*$ at 4.2K. We indeed observe a sign reversal of $T_0^*$ when cooling the sample below 1K, since the noise of the amplifier dominates at low enough temperature (see Figs. \ref{fighf} and \ref{fig_S3q}).  The non-linear behavior at low voltage at T=4.2K is similar to the one observed at room temperature with the previous setup, see Fig. \ref{sampleb}, i.e. when the noise emitted by the sample is larger than the noise emitted by the amplifier, revealing the contribution of the feedback of the environment.

In Fig. \ref{fig_S3q}a we present the result for $h\bar f<k_BT$ and $h\bar f>k_BT$. Those are obtained with an HF bandwidth being HF=4.5-5.5GHz (5GHz$\cong250$mK). The slope at high voltage has changed whereas it was temperature independent up to 290K. In the regime $h\bar f>k_BT$ the slope at high voltage is zero, as if only the term $\left< i_0i^2\right>$ remained. Further experiments are required to see if it is a coincidence for the particular values of $T$ or $f$. It is however clear that the result is different in the two regimes.

\subsection{Noise thermal impedance}

In the treatment of the effect of an external source of noise on the probability $P(i)$ we have supposed that $P$ responds instantaneously to the external voltage fluctuations. In fact one may ask how fast can $P(i)$ react. We exemplify this discussion in the case the second moment $S_{I^2}$ of a diffusive wire.

For a macroscopic wire, $S_{I^2}=4k_BTG$ is the equilibrium Johnson noise. A small voltage variation $\delta V\ll V$ induces a variation of the Joule power dissipated in the sample, $\delta P_J=2GV\delta V$ which in turn induces a variation of the temperature $\delta T=G_{th}^{-1}\delta P_J$. This will take place as long as the time scale of the variations is much smaller than the thermalization time of the wire. For a phonon-cooled wire ($L\gg L_{e-ph}$ with $L_{e-ph}$ the electron-phonon length) this time will be the electron-phonon interaction time. For a wire in the hot electron- diffusion cooled regime ($L_{e-ph}\gg L\gg L_{e-e}$ with $L_{e-e}$ the electron-electron inelastic length), the thermalization is obtained by hot electrons leaving the sample, which occurs at a time scale given by the diffusion time. For a short wire ($L_{e-e}\gg L$), no thermalization occurs within the electron gas. The noise emitted by the sample depends on the distribution function of the electrons in the wire, whose dynamics is diffusive. Therefore, the relevant time scale in this regime of elastic transport is again set by the diffusion time. Note that in this regime the temperature is no longer defined, neither is the thermal impedance. However, one can define a "noise thermal impedance" that extends the definition of the usual thermal impedance to cover any case \cite{NTI}.
The noise thermal impedance clearly determines the electrical effect of the environment at finite frequency. According to our simple calculation of the thermal effect of the environment, it also determines $S_{I^3}$ in the hot electron regime. The case of the elastic case is less clear, but should be qualitatively the same\cite{Pilgram_diff}.

\begin{figure}
\includegraphics[width= 0.8\columnwidth]{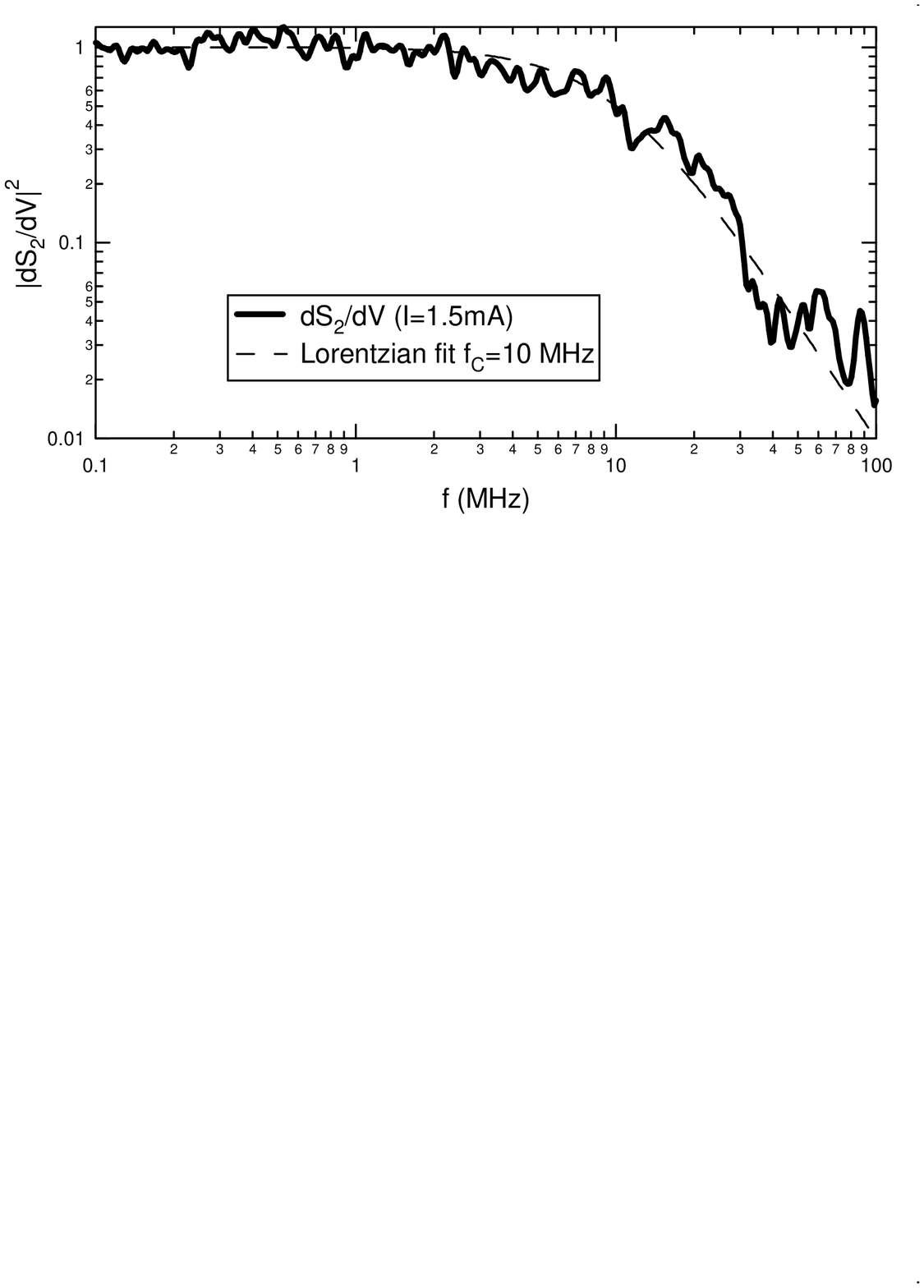}
\caption{$|dS_{V^2}/dV|^2$ renormalized to its value at low frequency, versus measurement frequency}
\label{fig_NTI}
\end{figure}

In order to illustrate the noise thermal impedance, we show a preliminary measurement of $dS_{V^2}/dV$ at high frequency in a $100\mu$m long diffusive Au wire, see Fig.\ref{fig_NTI}. The sample is biased by $I(t)=I_{dc}+\delta I\cos\omega t$ with $\delta I\ll I_{dc}$. $\omega$ is varied between 100kHz and 100MHz. With a setup similar to the one of Fig. \ref{fighf}a, we detect how much the rms amplitude of the noise, measured by the diode, oscillates at frequency $\omega$, which is similar to the measurement of $S_{V^3}$ in the presence of an external excitation. One clearly sees in the figure the cut-off at 10MHz, which probably corresponds to the electron-phonon time in the sample at $T=4.2K$. The curve can be well fitted by a Lorentzian, as expected, from which we deduce $\tau_{e-ph}=16$ns. Diffusion cooling would have led to a cut-off frequency of $\sim800$kHz. A study for different length at various temperatures is in progress.

\subsection{Conclusion}

In this article we have chosen to give simple pictures of more complicated phenomena. We hope this will encourage the reader to go through more detailed literature. We have shown experimental evidence of the existence of non-gaussian shot noise. As we have sketched in the last section, this is not the end of the story, and we hope to motivate the emergence of new experiments in this domain.

\section{Acknowledgements}

This work has been accomplished at Yale University in the group of Prof. D. Prober, who I warmly thank for his confidence and enthusiasm, as well as for having given me the possibility to perform this research. This was possible thanks to the flexibility of the CNRS, which allowed me go to Yale during four years. I acknowledge J. Senzier for his help at the beginning of this adventure. L. Spietz, C. Wilson and M. Shen are fondly thanked for providing me with samples. I also acknowledge fruitful discussions with many colleagues, among them C. Beenakker, W. Belzig, M. B\"uttiker, A. Clerk, M. Devoret, M. Kindermann, L. Levitov, Y. Nazarov, S. Pilgram and P. Samuelsson. This work was supported by NSF DMR-0072022 and 0407082.


\begin{thebibliography}{1}


\bibitem{BuBlan} Y.M. Blanter and M. B\"uttiker, Phys. Rep. \textbf{336}, 1 (2000).

\bibitem{LevitovMath} L.S. Levitov, H. Lee and G.B. Lesovik, J. Math. Phys. {\bf37}, 4845 (1996).
\bibitem{Chtche} G.B. Lesovik and N.M. Chtchelkatchev, JETP Lett. {\bf77}, 393 (2003).


\bibitem{Pilgram_diff} 
S. Pilgram, K. Nagaev and M. B\"uttiker, Phys. Rev. \textbf{B70} 045304 (2004).

\bibitem{Saleur} I. Safi and H. Saleur, Phys. Rev. Lett. \textbf{93}, 126602 (2004).
\bibitem{Zaikin_cumulants} 
A.V. Galaktionov, D.S. Golubev and A.D. Zaikin, Phys. Rev. \textbf{B68}, 085317 (2003).
\bibitem{Ingold} G.-L. Ingold and Yu. V. Nazarov in \textit{Single Charge Tunneling}, edited by H. Grabert and M. H. Devoret (Plenum Press, New York, 1992).
\bibitem{yeyati} A. Levy Yeyati, A. Martin-Rodero, D. Esteve and C. Urbina, Phys. Rev. Lett. \textbf{87} 046802 (2001).
\bibitem{gefen_tunnel} 
D.B. Gutman and Y. Gefen, Phys. Rev. \textbf{B64}, 205317 (2001).

\bibitem{Kindermann} M. Kindermann, Yu V. Nazarov and C.W.J. Beenakker, Phys. Rev. Lett. \textbf{90}, 246805 (2003)
\bibitem{Kindermann2}  C.W.J. Beenakker, M. Kindermann and Yu V. Nazarov, Phys. Rev. Lett. \textbf{90}, 176802 (2003).

\bibitem{S3nous} B. Reulet, J. Senzier and D.E. Prober, Phys. Rev. Lett. {\bf91}, 196601 (2003).
\bibitem{SPIE} B. Reulet \textit{et al.}. Proceedings of SPIE, \textit{Fluctuations and Noise in Materials}, 5469-33:244-56 (2004).




\bibitem{richard} R. Deblock, E. Onac, L. Gurevich and L.P. Kouwenhoven, Science {\bf301}, 203 (2003).
\bibitem{tobiska} J. Tobiska and Yu V. Nazarov, Phys. Rev. Lett. \textbf{93}, 106801 (2004).
\bibitem{finland} P.K. Lindell \textit{et al.}, Phys. Rev. Lett. \textbf{93}, 197002 (2004).

\bibitem{Zaikin_freq} 
A.V. Galaktionov, D.S. Golubev and A.D. Zaikin, Phys. Rev. \textbf{B68}, 235333 (2003).

\bibitem{NTI} B. Reulet and D.E. Prober, unpublished (cond-mat/0501397).




\end{thebibliography}
\end{document}